\documentclass[twocolumn]{revtex4}

\usepackage{amsmath,amsfonts}
\usepackage{amssymb,latexsym}
\usepackage{color}
\usepackage{graphicx}

\newcommand{\ii}{\mathrm{i}}
\newcommand{\ud}{\mathrm{d}}
\newcommand{\um}{\mathrm{m}}
\newcommand{\uD}{\mathrm{D}}
\newcommand{\uM}{\mathrm{M}}
\newcommand{\uE}{\mathrm{E}}

\begin{document}

\title{Viscous control of minimum uncertainty state in hydrodynamics}
\author{T.\ Koide}
\email{tomoikoide@gmail.com,koide@if.ufrj.br}
\affiliation{Instituto de F\'{\i}sica, Universidade Federal do Rio de Janeiro, C.P. 68528,
21941-972, Rio de Janeiro, RJ, Brazil}
\begin{abstract}
A minimum uncertainty state for position and momentum of a fluid element is obtained. We consider a general fluid described by the Navier-Stokes-Korteweg (NSK) equation, which reproduces the behaviors of a standard viscous fluid, a fluid with the capillary action and a quantum fluid, with the proper choice of parameters.
When the parameters of the NSK equation is adjusted to reproduce Madelung's hydrodynamic representation 
of the Schr\"{o}dinger equation, 
the uncertainty relation of a fluid element reproduces the Kennard and the Robertson-Schr\"{o}dinger inequalities in quantum mechanics.
The derived minimum uncertainty state is the generalization of the coherent state and 
its uncertainty is given by a function of the shear viscosity.  
The viscous uncertainty can be smaller than the inviscid minimum value when the shear viscosity is smaller than a critical value which is similar in magnitude to the Kovtun-Son-Starinets (KSS) bound. 
This uncertainty reflects the information of the fluctuating microscopic degrees of freedom in the fluid 
and will modify the standard hydrodynamic scenario, for example, in heavy-ion collisions.
\end{abstract}

\maketitle

\section{introduction}

The uncertainty relation is an important feature in quantum physics and its comprehension requires unceasing improvement \cite{heisenberg,ozawa,ozawa2,ozawa3,ozawa4}.
A similar relation was recently proposed  
in classical viscous hydrodynamics \cite{koide18,koide_review20}.
The derived relations describe the uncertainty associated with position and momentum of a fluid element.
Differently from the quantum-mechanical relations, the finite minimum uncertainty is induced by 
thermal fluctuations. 
Nevertheless, the hydrodynamic relations have the same structure as 
the quantum-mechanical ones.
When we apply the hydrodynamic uncertainty relations to 
Madelung's hydrodynamic representation of the Schr\"{o}dinger equation, 
the well-known Kennard and Robertson-Schr\"{o}dinger inequalities in quantum mechanics are reproduced. 
However the minimum uncertainty state for this viscous uncertainty relations has not been derived.

In this paper, we consider a general fluid described by the following differential equation,
\begin{eqnarray}
&&  (\partial_t + {\bf v} \cdot \nabla ) v^{i} 
= 
-\frac{1}{\uM} \partial_i V + 2\kappa \partial_i \frac{\nabla^2 \sqrt{\rho}}{\sqrt{\rho}}  \nonumber \\
&& 
-  \frac{1}{\uM \rho } \partial_i  \left\{ P 
-\left( \mu + \frac{\eta}{D} \right)  \left( \nabla \cdot {\bf v}  \right)
\right\} 
+
\frac{1}{\uM \rho}\sum_{j=1}^D \partial_j  \left( \eta  E^{ij}  \right) 
\, ,   \label{eqn:nsk} \nonumber \\
\end{eqnarray}
where ${\bf v}$, $V$, $P$, $\eta$ and $\mu$ are the velocity field, the external potential, the pressure, the shear viscosity and the second coefficient of viscosity, respectively. 
The number of the spatial dimension is denoted by $D$.
The traceless symmetric stress tensor is defined by   
\begin{equation}
E^{ij} = \frac{1}{2} \left( \partial_i v^j  + \partial_j v^i \right) - \frac{1}{D} \left( \nabla \cdot {\bf v} \right) \delta_{ij} \, .
\nonumber 
\end{equation}
Normally, hydrodynamics is described using the mass distribution. 
For the sake of comparison with quantum mechanics, however, we use the distribution of constituent particles of the fluid $\rho$, which is normalized by the number of constituent particles $N$.
The mass distribution is given by $\uM \rho$ with $\uM$ being the mass of constituent particles of a simple fluid.
This equation is reduced to the Navier-Stokes-Fourier (NSF) equation when 
the last term on the first line is dropped.
We call this additional term the $\kappa$ term.

Equation (\ref{eqn:nsk}) appears at least in three applications of hydrodynamics. 
Korteweg considered that the behavior of liquid-vapor fluids near phase transitions is described by 
a generalized equation of fluid. 
This is called the Navier-Stokes-Korteweg (NSK) equation and Eq.\ (\ref{eqn:nsk}) is 
a special case of the NSK equation. 
Then the $\kappa$ term describes the capillary action \cite{korteweg}.
Brenner pointed out that, since the velocity of a tracer particle of a fluid is not necessarily parallel to 
the mass velocity, the existence of these two velocities should be taken into account 
in the formulation of hydrodynamics.
This theory is called bivelocity hydrodynamics \cite{koide18,koide_review20,brenner,gustavo,dadzie} 
and the NSK equation is understood to be one of the variants.
Lastly, the $\kappa$ term is equivalent to the so-called gradient of the quantum potential. 
Indeed the NSK equation becomes Madelung's hydrodynamics 
when we choose $\kappa = \hbar^2/(4\uM^2)$ in the vanishing viscosity limit.
In addition, Eq.\ (\ref{eqn:nsk}) is sometimes used as a model of a quantum viscous fluid \cite{brull2010,bresch19}.

The purpose of this paper is to derive the minimum uncertainty state of the fluid described by the NSK equation. 
As will be seen later, Eq.\ (\ref{eqn:nsk}) is formulated in the framework of 
the generalized variational principle, the stochastic variational method (SVM) \cite{yasue,zambrini,koide18,koide_review20,koide12,koide-review1,koide-review2,koide19,koide20-1}.
Then the uncertainty relation of the NSK fluid is derived by 
applying the method developed in Refs.\ \cite{koide18,koide20-1,koide_review20}.
We show that the minimum uncertainty state of the derived uncertainty relation 
is given by a generalized coherent state. 
We further find that this minimum uncertainty is controlled by the shear viscosity 
and can be smaller than the inviscid minimum value for sufficiently weak viscosity.
This uncertainty reflects the information of the fluctuating microscopic degrees of freedom in the fluid 
and will modify the standard hydrodynamic scenario, for example, in heavy-ion collisions.

This paper is organized as follows.
In Sec.\ \ref{sec:svm}, the NSK equation is formulated in the framework of the stochastic variational method \cite{yasue,zambrini,koide18,koide_review20,koide12,koide-review1,koide-review2,koide19,koide20-1}.
In Sec.\ \ref{sec:ucr}, the uncertainty relation is derived by applying the method in Refs.\ \cite{koide18,koide_review20}. 
In Sec.\ \ref{sec:mus}, we derive the minimum uncertainty state of the NSK equation and study the properties.
Concluding remarks and the possible influence in heavy-ion collision physics are discussed in Sec.\ \ref{sec:conclusion}.

\section{Stochastic variational method} \label{sec:svm}

To define the uncertainty relation in fluids, we formulate Eq.\ (\ref{eqn:nsk}) in 
SVM \cite{yasue,zambrini,koide18,koide_review20,koide12,koide-review1,koide-review2,koide19,koide20-1}.
As a similar but different approach, see Ref.\ \cite{kuipers}
As is well-known, the behavior of a fluid can be described by the ensemble of fluid elements.
We thus consider the variation of the trajectory of a fluid element in SVM.
A fluid element is an abstract volume element with a fixed mass 
and constituent particles inside of it are assumed to be thermally equilibrated.
For the sake of simplicity, however, we identify a fluid element with a constituent particle in the following discussion. 
See Ref.\ \cite{koide_review20} for more details on the uncertainty relation for fluid elements.

In SVM, the viscous and $\kappa$ terms are induced through the fluctuations of constituent particles (fluid elements).
Then the trajectory of a constituent particle is supposed to be given by the forward stochastic differential equation (SDE), 
\begin{eqnarray}
\ud \widehat{\bf r}(t) =  {\bf u}_+ (\widehat{\bf r}(t), t) \ud t + \sqrt{2\nu} \ud\widehat{\bf W}(t)  \, \, \, (\ud t > 0) \, . 
\nonumber 
\end{eqnarray} 
The second term on the right-hand side represents the noise of Brownian motion. 
We used $(\widehat{\,\,\,\,})$ to denote stochastic variables and $\ud \widehat{A}(t) = \widehat{A}(t+\ud t) - \widehat{A}(t)$ for an arbitrary $\widehat{A}(t)$. 
The standard Wiener process is described by $\widehat{\bf W}(t)$ which satisfies   
\begin{equation}
\begin{split}
\uE[ \ud\widehat{\bf W}(t) ] = 0\, , \, \, \, \, 
\uE [ \ud\widehat{W}^{i} (t) \ud\widehat{W}^{j} (t^\prime) ] = |\ud t| \, \delta_{t \,t^\prime} \, \delta_{ij}  \, ,
\end{split}
\label{eqn:wiener}
\end{equation}
where 
$\uE[\, \, \, \,]$ denotes the ensemble average for the Wiener process.
Note that ${\bf u}_+ (\widehat{\bf r}(t),t)$ is stochastic because of $\widehat{\bf r}(t)$, 
but ${\bf u}_+ ({\bf x},t)$ is a smooth function.
The field ${\bf u}_+ ({\bf x},t)$ is associated with the velocity of constituent particles.
The purpose of SVM is to determine its form by applying the variational principle.

The noise intensity $\nu$ controls the stochasticity of the trajectory. 
In the derivation of Madelung's hydrodynamics, 
$\nu$ is given by the function of the Planck constant \cite{yasue}.
In the derivation of the NSF equation, however, $\nu$ characterizes the intensity of thermal fluctuations 
and thus is a function of temperature \cite{koide12}. 
In this work, we consider that $\nu$ is a general function 
of the Planck constant and temperature.

The standard definition of velocity is not applicable in stochastic trajectories 
because the left and right-hand limits of the inclination of stochastic trajectories do not agree. To distinguish this difference, 
we consider the backward time evolution of the trajectory described by the backward SDE,  
\begin{eqnarray}
\ud \widehat{\bf r}(t) =  {\bf u}_- (\widehat{\bf r}(t), t) \ud t + \sqrt{2\nu} \ud \underline{\widehat{\bf W}}(t)  \, \, \, (\ud t < 0) \, , \nonumber
\end{eqnarray}
where $\ud \underline{\widehat{\bf W}}(t)$ satisfies the same correlation properties as Eq.\ (\ref{eqn:wiener}) using $|\ud t | = - \ud t$. 
The field ${\bf u}_- ({\bf x},t)$ is associated with the velocity backward in time.

Because of this ambiguity of velocity, 
Nelson introduced two different time derivatives \cite{nelson}: 
one is the mean forward derivative $\uD_+$ and the other the mean backward derivative $\uD_-$,
which are defined by 
\begin{equation}
\uD_\pm   \widehat{\bf r}(t)  = \lim_{\ud t \rightarrow0\pm} \uE \left[  \frac{ \widehat{\bf r}({t + \ud t}) -
\widehat{\bf r}(t)}{\ud t} \Big| \widehat{\bf r}(t) \right] = {\bf u}_\pm (\widehat{\bf r}(t), t) \,  . \label{eqn:mfd}
\end{equation}
Here the expectation value is the conditional average for fixing $\widehat{\bf r}(t)$ and we used that $\widehat{\bf r}(t)$ is Markovian. 
When these are operated to a function of $\widehat{\bf r}(t)$, 
we find
\begin{eqnarray}
 \uD_\pm f(\widehat{\bf r}(t),t) \Big|_{\widehat{\bf r}(t) = {\bf x}} 
\hspace{-0.1cm}= 
\left\{
\partial_t  + {\bf u}_\pm ({\bf x},t) \cdot \nabla \pm \nu \nabla^2
\right\} f({\bf x},t) \, , \label{eqn:ito}
\end{eqnarray} 
where $f({\bf x},t)$ is an arbitrary smooth function and we used Ito's lemma \cite{koide_review20}.
That is, $\uD_+$ and $\uD_-$ correspond to material derivatives along the stochastic trajectories described by 
the forward and backward SDE's, respectively.
These derivatives satisfy the following relation,
\begin{eqnarray}
\lefteqn{\int^{t_b}_{t_a} \ud t \, \uE \left[
\widehat{B}(t) \uD_+ \widehat{A}(t) + \widehat{A}(t) \uD_- \widehat{B}(t) 
\right]} && \nonumber \\
&&= \uE \left[
\widehat{A}(t_b) \widehat{B}(t_b) - \widehat{A}(t_a) \widehat{B}(t_a)
\right] \, . \nonumber 
\end{eqnarray}
This corresponds to the stochastic generalization of integration by parts \cite{koide_review20}.

The particle distribution is defined by 
\begin{eqnarray}
\rho ({\bf x},t) = \int \ud^D {\bf R} \, \rho_0 ({\bf R}) \uE[\delta({\bf x} - \widehat{\bf r}(t))] \, , \nonumber
\end{eqnarray}
where ${\bf R} $ denotes the initial position of the constituent particles and 
its distribution is characterized by $\rho_0 ({\bf R})$.
Applying the forward and backward SDE's to this definition, 
two Fokker-Planck equations are obtained,
\begin{eqnarray}
\partial_t \rho({\bf x},t) &=& -\nabla \cdot \left\{ {\bf u}_+ ({\bf x},t) \rho({\bf x},t) \right\}+ \nu \nabla^2 \rho({\bf x},t) \, , \label{eqn:ffp}\\
\partial_t \rho({\bf x},t) &=& -\nabla \cdot \left\{ {\bf u}_- ({\bf x},t) \rho({\bf x},t) \right\}- \nu \nabla^2 \rho({\bf x},t) \, .  \label{eqn:bfp}
\end{eqnarray}
The first and second equations are obtained using the forward and backward SDE's, respectively. 
The different sign in the second terms on the right-hand sides is due to $|\ud t|$ in the correlation function of the Wiener process (\ref{eqn:wiener}).
That is, the second term of Eq.\ (\ref{eqn:ffp}) represents the diffusion effect induced by the noise term, but 
the corresponding term in Eq.\ (\ref{eqn:bfp}) gives the accumulation effect.
These equations should be equivalent.
To conform Eq.\ (\ref{eqn:bfp}) to Eq.\ (\ref{eqn:ffp}), ${\bf u}_- ({\bf x},t)$ should be chosen to 
satisfy the consistency condition,
\begin{eqnarray}
 {\bf u}_+ ({\bf x},t) =  {\bf u}_- ({\bf x},t) + 2\nu \nabla \ln \rho ({\bf x},t) \label{eqn:cc}
\, .
\end{eqnarray}
See Ref.\ \cite{koide_review20} for details.
It is also noteworthy that a similar condition plays an important role in bivelocity hydrodynamics \cite{koide18,koide_review20,brenner,gustavo,dadzie}.

Let us consider the classical Lagrangian,
\begin{eqnarray}
L_{cla} ({\bf r}, \ud {\bf r}/\ud t ) =\frac{\uM}{2} \left( \frac{\ud {\bf r}}{\ud t} \right)^2 - V  - \frac{\varepsilon}{\rho} \, , \label{eqn:cla-lag}
\end{eqnarray}
where $\varepsilon$ is an internal energy density given by a function of the particle distribution and the entropy density.
Applying the classical variation, this Lagrangian gives ideal-fluid dynamics (Euler equation) \cite{koide_review20}.
As mentioned before, the viscous and $\kappa$ terms are induced through the fluctuating trajectory in SVM and hence the NSK equation (\ref{eqn:nsk}) is obtained by applying SVM to this Lagrangian.
To find the corresponding stochastic Lagrangian,
we have to replace $\ud/\ud t$ with $\uD_+$ and $\uD_-$ in Eq.\ (\ref{eqn:cla-lag}). 
Due to this ambiguity in the replacement, we introduce two real parameters $\alpha_A$ and $\alpha_B$.
Then the stochastic Lagrangian is defined by
\begin{eqnarray}
\lefteqn{L_{sto} (\widehat{\bf r},\uD_+ \widehat{\bf r}, \uD_- \widehat{\bf r}) } && \nonumber \\
&& = 
\frac{\uM}{2} (\uD_+ \widehat{\bf r}, \uD_- \widehat{\bf r}) 
{\cal M} 
\left(
\begin{array}{c}
\uD_+ \widehat{\bf r} \\
\uD_- \widehat{\bf r}
\end{array}
\right) 
 - V
- \frac{\varepsilon}{\rho} \, , 
 \label{eqn:sto-lag} 
\end{eqnarray}
with
\begin{eqnarray}
{\cal M} 
= 
\left(
\begin{array}{cc}
\left( \frac{1}{2} + \alpha_A \right)\left( \frac{1}{2} + \alpha_B \right)
& \frac{1}{4} - \frac{\alpha_B}{2}  \\
\frac{1}{4} - \frac{\alpha_B}{2}  & \left( \frac{1}{2} - \alpha_A \right)\left( \frac{1}{2} + \alpha_B \right)
\end{array}
\right) \, . \nonumber
\end{eqnarray}
See the discussion in Sec.\ 4.1 in Ref.\ \cite{koide_review20} for details.
In the vanishing limit of $\nu$,   
$\uD_\pm$ coincide with $\ud/\ud t$ 
and then the stochastic Lagrangian  (\ref{eqn:sto-lag}) is reduced to 
the corresponding classical one (\ref{eqn:cla-lag}) independently of $\alpha_A$ and $\alpha_B$.
The parameters $\alpha_A$ and $\alpha_B$ are absorbed into the definitions of $\kappa$ and $\eta$  
as shown later in Eq.\ (\ref{eqn:coeff}).

In the classical variation, a trajectory is entirely determined for a given velocity. 
This is however not the case with SVM due to the noise terms  
in the two SDE's. 
Therefore only the averaged behavior of the stochastic Lagrangian is optimized by variation. 
The action is then defined by the expectation value, 
\begin{eqnarray}
I_{sto} [\widehat{\bf r}] 
= \int^{t_f}_{t_i} \ud t \, \uE[L_{sto}(\widehat{\bf r},\uD_+ \widehat{\bf r}, \uD_- \widehat{\bf r})]\, ,
\label{eqn:sto_act}
\end{eqnarray}
with an initial time $t_i$ and a final time $t_f$. 
Here, the initial distribution of constituent particles $\rho_0 ({\bf R})$ is omitted but it does not affect the result of the stochastic variation. 
See, for example, Eq.\ (116) in Ref.\ \cite{koide_review20}.

The variation of the stochastic trajectory is defined by  
$\widehat{\bf r}(t) \longrightarrow \widehat{\bf r}^\prime (t) = \widehat{\bf r}(t) + \delta {\bf f} (\widehat{\bf r}(t),t)$, where 
an infinitesimal smooth function $\delta {\bf f} ({\bf x},t)$ satisfies $\delta {\bf f}({\bf x},t_i) = \delta {\bf f}({\bf x},t_f) = 0$. 
We further define the fluid velocity field by  
\begin{eqnarray}
{\bf v} ({\bf x},t) = \frac{{\bf u}_+ ({\bf x},t) + {\bf u}_- ({\bf x},t)}{2} \, . \label{eqn:def_v}
\end{eqnarray}
Then the stochastic variation of Eq.\ (\ref{eqn:sto_act}) leads to 
\begin{eqnarray}
\left[
\frac{\uD_- {\bf p}_+ + \uD_+ {\bf p}_-}{2}
 = - \nabla V  -  \frac{1}{ \rho } \nabla  \left\{ P  -  \mu (\nabla \cdot {\bf v}  ) \right\} 
\right]_{\widehat{\bf r}(t) = {\bf x}}
\, . \label{eqn:qvh_nn}
\end{eqnarray}
Here, $\mu$ is obtained through the variation of the entropy density. See Sec. 5.1 in Ref.\ \cite{koide12} for details.
To obtain $P$, $\varepsilon$ is assumed to satisfy the local thermal equilibrium in the variation of Eq.\ (\ref{eqn:sto_act}).
See the discussion around Eq.\ (106) in Ref.\ \cite{koide_review20}.
The $\varepsilon$ and potential terms, however, do not affect the definitions of the two momenta, which are introduced through the Legendre transformation of the stochastic Lagrangian, 
\begin{eqnarray}
{\bf p}_\pm  ({\bf x},t ) = \left. 2\frac{\partial L_{sto}}{\partial (\uD_\pm \widehat{\bf r})} \right|_{\widehat{\bf r} = {\bf x}}  \, .
\label{eqn:twomom}
\end{eqnarray}
Here the factors $2$ in the definitions of ${\bf p}_\pm ({\bf x},t )$ are introduced for a convention to reproduce the classical result in the vanishing limit of $\nu$ \cite{koide18}. 
Note that the operations of $\uD_\pm$ to ${\bf p}_\mp ({\bf x},t )$ are calculated using Eq.\ (\ref{eqn:ito}).
Then it is straightforward to show that Eq.\ (\ref{eqn:qvh_nn}) reproduces the NSK equation (\ref{eqn:nsk}) with the identification, 
\begin{eqnarray}
\begin{split}
\kappa = 2 \alpha_B \nu^2 \, , \, \, \, \, 
\eta = 2\alpha_A (1 + 2 \alpha_B) \nu \uM \rho \, .
\end{split}
\label{eqn:coeff}
\end{eqnarray}

Using  ${\bf v} ({\bf x},t)$ defined by Eq.\ (\ref{eqn:def_v}), the two Fokker-Planck equations (\ref{eqn:ffp}) and (\ref{eqn:bfp})
are simplified and the equation of continuity is obtained,
\begin{eqnarray}
\partial_t \rho ({\bf x},t) + \nabla \cdot ( \rho({\bf x},t) {\bf v}({\bf x},t) ) =0 \, . \nonumber
\end{eqnarray}

It is important to note that 
the NSK equation (\ref{eqn:nsk}) reproduces not only the Schr\"{o}dinger equation but also the Gross-Pitaevskii equation when we choose the internal energy density $\varepsilon$ and the parameters in the stochastic Lagrangian $(\alpha_A,\alpha_B)$ appropriately.  
See the discussion in Refs.\ \cite{koide18,koide_review20,koide12} for details.

\section{Uncertainty relations} \label{sec:ucr}

The emergence of the two momenta is attributed to the non-differentiability of 
the stochastic trajectory.
As seen in Eq.\ (\ref{eqn:qvh_nn}), ${\bf p}_\pm ({\bf x},t )$ contribute to our equation of motion on an equal footing. Therefore it is natural to define the standard deviation of momentum by the average of the two contributions, 
${\bf p}_+ ({\bf x},t )$ and ${\bf p}_- ({\bf x},t )$.

We define the standard deviations of position and momentum. 
The former is given by 
\begin{eqnarray}
\sigma^{(2)}_{x^{i}}  =  \lceil (\delta {x}^{i} )^2 \rfloor \, , \nonumber
\end{eqnarray}
where $\delta {f} =  f ({\bf x},t) - \lceil {f} \, \rfloor$ and 
we introduced the following expectation value,
\begin{eqnarray}
\lceil f \, \rfloor 
= 
\frac{1}{N}\int \ud^D {\bf x} \, \rho ({\bf x},t) f ({\bf x},t)
\, , \label{eqn:exp}
\end{eqnarray} 
with $N$ being the number of constituent particles. 
As discussed above, the latter is given by  
the average,
\begin{eqnarray}
 \sigma^{(2)}_{p^{i}}
&=& \frac{\lceil (\delta {p}^{i}_+ )^2 \rfloor
+ \lceil (\delta {p}^{i}_-  )^2 \rfloor  }{2} \nonumber \\
&=&
\left\lceil \left( \frac{\delta p^i_- + \delta p^i_+ }{2} \right)^2 \right\rfloor 
+ \left\lceil \left( \frac{\delta p^i_- - \delta p^i_+ }{2} \right)^2 \right\rfloor 
\, , \nonumber
\end{eqnarray}
where
\begin{eqnarray}
\left(
\begin{array}{c}
{\bf p}_-({\bf x},t) - {\bf p}_+ ({\bf x},t) \\
{\bf p}_-({\bf x},t) + {\bf p}_+ ({\bf x},t) 
\end{array}
\right) 
=
2\uM G
\left(
\begin{array}{c}
-\nu \nabla \ln \rho ({\bf x},t) \\ 
{\bf v} ({\bf x},t) 
\end{array}
\right)
 . \label{eqn:p-pp+p}
\end{eqnarray}
The symmetric matrix $G$ is defined by 
\begin{eqnarray}
G = 
\left(
\begin{array}{cc}
\frac{\kappa}{\nu^2} & - \frac{\xi}{\nu} \\
- \frac{\xi}{\nu} & 1
\end{array}
\right) \, , 
\nonumber
\end{eqnarray}
with the kinematic viscosity, 
\begin{eqnarray}
\xi = \frac{\eta}{ 2\uM \rho} \, .
\nonumber
\end{eqnarray}
The consistency condition (\ref{eqn:cc}) is used in Eq.\ (\ref{eqn:p-pp+p}).
The above definitions of the standard deviations reproduce the corresponding quantum-mechanical quantities 
as shown later in Eq.\ (\ref{eqn:op-sto}).

Using these definitions and the Cauchy-Schwarz inequality $\lceil A^2 \rfloor \lceil B^2 \rfloor \ge (\lceil AB \rfloor)^2$, 
the product of $\sigma^{(2)}_{x^i}$ and $\sigma^{(2)}_{p^j}$ is shown to satisfy the inequality, 
\begin{eqnarray}
\lefteqn{\sigma^{(2)}_{x^i} \sigma^{(2)}_{p^j} } && \nonumber \\
&\ge& 
\uM^2
\left[ 
\frac{\nu^2 \lambda^2_+ \lambda^2_-}{\lambda_+ + \lambda_- - \lambda_+ \lambda_- }
 \delta_{ij}  + 
 \left(
\lambda_+ + \lambda_- - \lambda_+ \lambda_-
\right) \right. \nonumber \\
&& \left. \times \left\{ \lceil \delta x^i \delta v^j \rfloor 
+ \frac{\nu^2}{\xi} 
 \frac{(\lambda_+ + \lambda_-) (1-\lambda_+)(1-\lambda_-)}{\lambda_+ + \lambda_- - \lambda_+ \lambda_-} 
 \delta_{ij} \right\}^2 \right] \nonumber \\
&=&
\uM^2 
\frac{(\xi^2 - \kappa)^2}{\nu^2 + \xi^2}  \delta_{i j}  \nonumber \\
&&  + \uM^2 \left( 1+ \frac{\xi^2}{\nu^2}\right) 
\left(
\lceil \delta x^i \delta v^j \rfloor - \frac{\xi (\nu^2+\kappa)}{\nu^2 + \xi^2} \delta_{i j}
\right)^2 \, ,
\label{eqn:ucr}
\end{eqnarray}
where $\lambda_\pm = \{ 1+ \kappa/\nu^2 \pm \sqrt{(1-\kappa/\nu^2)^2 + 4 \xi^2/\nu^2} \}/2$ are the eigenvalues of $G$.
This inequality was derived in Ref. \cite{koide18} for the first time. 
The right-hand side becomes minimum when 
$\lceil \delta x^i \delta v^j \rfloor = \delta_{i j} \xi (\nu^2+\kappa)/(\nu^2 + \xi^2) $.

The inequality reproduces the well-known result in quantum mechanics by choosing  
\begin{eqnarray}
(\alpha_A,\alpha_B,\nu) = \left( 0, \frac{1}{2}, \frac{\hbar}{2\uM} \right) \, . \label{eqn:para_qm}
\end{eqnarray}
Then Eq.\ (\ref{eqn:nsk}) (or equivalently Eq.\ (\ref{eqn:qvh_nn})) coincides with Medelung's hydrodynamics, and  
our uncertainty relation (\ref{eqn:ucr}) leads to the Robertson-Schr\"{o}dinger inequality, 
\begin{eqnarray}
\sigma^{(2)}_{x^i} \sigma^{(2)}_{p^j}
\ge
\frac{\hbar^2}{4} \delta_{ij} + \left\{ {\rm Re} [ \langle (x^i_{op} - \langle x^i_{op}) (p^j_{op} - \langle p^j_{op}) \rangle ] \right\}^2 \, . \label{eqn:rs_qm}
\end{eqnarray} 
In this derivation, we used that 
the quantum-mechanical expectation values are expressed as  
\begin{eqnarray}
\begin{array}{lcl}
 \langle {x}^i_{op} \rangle = \lceil x^i \rfloor \, ,  & & \langle ({x}^i_{op} - \langle x^i_{op} \rangle)^2 \rangle =  \sigma^{(2)}_{x^i}  \, ,\\
\langle {p}^i_{op} \rangle  = \frac{\lceil p^i_+ \rfloor + \lceil p^i_- \rfloor}{2} \, , & & \langle ({p}^i_{op} -\langle p^i_{op} \rangle)^2 \rangle = \sigma^{(2)}_{p^i}  \, ,
\end{array}
\label{eqn:op-sto}
\end{eqnarray}
where ${\bf x}_{op}$ and ${\bf p}_{op}$ are the position and momentum operators, respectively, and 
$\langle \, \, \, \, \rangle$ denotes the expectation value with a wave function. 
See Refs.\ \cite{koide18,koide_review20} for details.
The second term on the right-hand side of Eq.\ (\ref{eqn:rs_qm}) is always positive.
The Kennard inequality is reproduced when this term is ignored.

Note that the famous paradox for the angular uncertainty relation is resolved in the present approach \cite{koide20-1}.
For a quantum-mechanical uncertainty relation in 
different stochastic approaches, see Refs.\ \cite{illuminati,lindgren}. 
The advantage of the present approach compared to the standard operator formalism is discussed in Sec.\ \ref{sec:conclusion}.

\subsection{Zero uncertainty} \label{sec:ideal_limit}

The NSK equation is reduced to the Euler equation in the vanishing noise limit 
$\nu \rightarrow 0$.
Then our inequality (\ref{eqn:ucr}) becomes
\begin{eqnarray}
\sigma^{(2)}_{x^i} \sigma^{(2)}_{p^j} \ge 0 \, . \nonumber
\end{eqnarray}
Here we dropped the second term on the right-hand side of  Eq.\ (\ref{eqn:ucr}) to find the Kennard-type inequality.

One may consider that the zero uncertainty can be realized even for fluctuating dynamics 
by setting $\kappa = \xi^2$.
This choice of the parameter is however not permitted. 
It is because the right-hand side of Eq.\ (\ref{eqn:ucr}) can be reexpressed as 
\begin{eqnarray}
\sigma^{(2)}_{x^i} \sigma^{(2)}_{p^j} 
&\ge& \frac{(4 \uM \nu)^2}{1+(\xi/\nu)^2} |{\rm det} ({\cal M}) |^2 \delta_{ij} \, . \nonumber 
\end{eqnarray}
Here, again, the irrelevant second term on the right-hand side of Eq.\ (\ref{eqn:ucr}) is ignored. 
The matrix ${\cal M}$ is introduced in Eq.\ (\ref{eqn:sto-lag}).
That is, the condition $\kappa = \xi^2$ is equivalent to ${\rm det} ({\cal M}) =0$. 
However ${\rm det} ({\cal M})$ cannot disappear to define our momenta 
through the Legendre transformation of the stochastic Lagrangian.  
Therefore the uncertainty for stochastic dynamics always has a finite value.

\section{Viscous minimum uncertainty state}
\label{sec:mus}

We discuss the minimum uncertainty state of the inequality (\ref{eqn:ucr}) in one-dimensional system.
Such a state should reproduce the well-known result in quantum mechanics when we choose Eq.\ (\ref{eqn:para_qm}).
To guess the viscous minimum uncertainty state, we consult the numerical result of the NSF equation.
The time evolutions of the uncertainty of the NSF fluid are numerically calculated in Ref.\ \cite{koide_review20} 
using the initially Gaussian distribution of the mass at rest.
As shown in Fig.\ 7, the uncertainty of the viscous fluids with low Reynolds numbers takes a value close to the theoretically predicted minimum soon after the initial time.
The profiles of the mass distribution and the velocity field are shown in Fig.\ 4.
The mass distribution is then approximately given by a Gaussian function.
Ignoring the behavior in the irrelevant low density region, the velocity field seems to have a linear position dependence.
We thus assume that the viscous minimum uncertainty state is given by  
\begin{eqnarray}
\begin{split}
\rho(x) = \sqrt{\frac{A}{\pi}} e^{-A (x-x_0)^2} \, , \, \,\,\, v(x) = v_0 + B x \, ,
\end{split}
\label{eqn:mu-state}
\end{eqnarray}
where $x_0$, $v_0$, $A$ and $B$ are real constants. 
More properly, the second equation should be $v(x) = v_0 + B (x - x_0)$ but the difference is absorbed into the definition of $v_0$.
The standard deviations, $\sigma^{(2)}_x$ and $\sigma^{(2)}_p $, are easily calculated using this state and then 
we find
\begin{eqnarray}
\sigma^{(2)}_x \sigma^{(2)}_p 
&=& \uM^2 \frac{(\kappa - \xi^2)^2 }{\nu^2 + \xi^2} \nonumber \\
&& + \uM^2 \left( 1 + \frac{\xi^2}{\nu^2} \right)
\left( \frac{B}{2A} -  \frac{\xi(\nu^2 + \kappa)}{\nu^2 + \xi^2} \right)^2 
\, .
\label{eqn:minim_product}
\end{eqnarray}
Therefore, by choosing  
\begin{eqnarray}
\frac{B}{2A} = \lceil \delta x \delta v \rfloor  =  \frac{\xi(\nu^2 + \kappa)}{\nu^2 + \xi^2} \, , \label{eqn:B}
\end{eqnarray}
we find that Eq.\ (\ref{eqn:minim_product}) gives the minimum of  
the inequality (\ref{eqn:ucr}).
That is, the viscous minimum uncertainty state is defined by Eqs.\ (\ref{eqn:mu-state}) and (\ref{eqn:B}).

Because of the position-dependent velocity, the viscous minimum uncertainty state expands to homogenize the particle distribution.
Therefore the lifetime of the viscous minimum uncertainty state will be short in general.
To see the influence of the viscous uncertainty, we should observe short time evolutions in small inhomogeneous systems as is realized 
in heavy-ion collisions. See also the discussion in Sec.\ \ref{sec:conclusion}.

\subsection{NSF equation} \label{sec:nsf}

The above result describes the minimum uncertainty state for the NSF equation by setting $\kappa=0$,
\begin{eqnarray}
\rho (x) = \sqrt{\frac{A}{\pi}} e^{-A (x-x_0)^2} \, , \, \,\,\, v(x) = v_0 + 2A \frac{\xi \nu^2}{\nu^2 + \xi^2} x \, . \nonumber 
\end{eqnarray}
As pointed out, the linear-position dependence of the velocity field is observed in the expansion process of a localized fluid which is shown 
in Fig.\ 4 of Ref.\ \cite{koide_review20}.
As the initial condition, we used the stationary fluid given by the Gaussian mass distribution.
The isentropic ideal gas is considered for the equation of state.

The corresponding minimum value is 
\begin{eqnarray}
\sigma^{(2)}_x \sigma^{(2)}_p 
&=& \uM^2 \frac{ \xi^4 }{\nu^2 + \xi^2} \, . \nonumber 
\end{eqnarray}
The minimum value is characterized by two different parameters and thus, 
even if we fix $\xi$, the minimum value is affected by $\nu$.
Because the diffusion coefficient of the Fokker-Planck equations obtained from the SDE's is given by $\nu$, 
it is natural to consider that the noise intensity $\nu$ is determined by the diffusion coefficient of a fluid.

For example, let us consider water \cite{koide_review20}. The mass of a water molecule is $\uM = 3 \times 10^{-26}$ kg. 
At room temperature, the kinematic viscosity is $\xi \sim 10^{-6}$ $\um^2/s$ and the diffusion coefficient in the liquid phase is $\nu \sim 10^{-9}$ $\um^2/s$.
Thus, the contribution of $\nu$ is negligibly smaller than $\xi$ and the minimum value is given by 
\begin{eqnarray}
\uM^2 \frac{ \xi^4 }{\nu^2 + \xi^2} \sim \uM^2 \xi^2 \sim \, 600 \times \frac{\hbar}{2} \, . \nonumber 
\end{eqnarray}
For water vapor, $\xi \sim 0.3 \times 10^{-6}$ $\um^2/s$ and $\nu \sim 10^{-4}$ $\um^2/s$.
Contrary to the above case of liquid, the effect of diffusion is larger than the viscosity in the gas phase and 
the minimum value becomes 
\begin{eqnarray}
\uM^2 \frac{ \xi^4 }{\nu^2 + \xi^2} \sim \uM^2 \frac{\xi^4}{\nu^2} \sim \, 60 \times \frac{\hbar}{2} \, . \nonumber
\end{eqnarray}
One can see that these minimum values are much larger than the corresponding quantity in quantum mechanics.
These are however still much smaller than the coarse-graining scale in the standard applications of hydrodynamics 
and thus this effect will be irrelevant to most of applications. 
See also the discussion in Sec.\ \ref{sec:conclusion}.

The above results suggest that 
the difference of liquid and gas can be characterized by the different behaviors of the uncertainty.
See the discussion in Ref.\ \cite{koide_review20} for details.

\subsection{Quantum-mechanical limit} \label{sec:qm_limit}

The viscous minimum uncertainty state is the generalization of the (standard) coherent state. 
In Madelung's hydrodynamics \cite{madelung}, 
$\rho (x)$ and $v(x)$ for a given wave function $\Psi(x)$  
are defined by 
\begin{eqnarray}
\begin{split}
\rho (x) = |\Psi(x)|^2 \, , \, \, \, \, 
v (x) = -\ii \frac{\hbar}{\uM} {\rm Im}[ \partial_x \ln \Psi(x)] \, .
\end{split}
\label{eqn:decomp}
\end{eqnarray}
At the same time, the coordinate representation of the coherent state is given by
\begin{eqnarray}
\langle x | \alpha \rangle = \left( \frac{C}{\pi}\right)^{1/4} e^{-\frac{1}{2}(\sqrt{C}x- \alpha_R)^2 } 
e^{\ii \alpha_I (\sqrt{C} x - \alpha_R/2)} \, ,
\nonumber
\end{eqnarray}
where $\alpha_R$, $\alpha_I$ and $C$ are real constants and $\alpha = (\alpha_R + \ii \alpha_I)/\sqrt{2}$ is the eigenvalue of the lowering operator in 
a quantum harmonic oscillator \cite{book:jpg}.
Substituting this into Eq.\ (\ref{eqn:decomp}), we find that $\rho(x)$ and $v(x)$ for the coherent state are reproduced from Eq.\ (\ref{eqn:mu-state})
by choosing 
\begin{eqnarray}
\begin{array}{lcl}
A = C \, , & & x_0 = \frac{\alpha_R}{\sqrt{C}} \, ,\\
B = 0\, ,& & v_0 = \frac{\hbar}{\uM} \sqrt{C} \alpha_I \, .
\end{array}
\nonumber 
\end{eqnarray}
Here $B=0$ means the absence of the kinematic viscosity in Eq.\ (\ref{eqn:B}).

Note that this state becomes stationary and gives the ground state of the harmonic oscillator in quantum mechanics 
when the parameters are chosen by 
\begin{eqnarray}
C = \frac{\uM \omega}{\hbar} \, , \, \, \, \, \, \, \alpha_R = \alpha_I = 0\, 
\end{eqnarray}
where $\omega$ is the angular frequency of the harmonic potential.
Indeed, by using the definition of $v(x)$ (\ref{eqn:def_v}) and the consistency condition (\ref{eqn:cc}), 
we can show that this state gives the stationary point of the Fokker-Planck equations (\ref{eqn:ffp}) and (\ref{eqn:bfp}). 
See also the discussion in Sec.\ \ref{sec:smv}.

In the viscous case where $B \neq 0$, the minimum uncertainty state is not stationary.
In other words, there exist other states which have smaller uncertainties around the 1-D Gaussian fluid with $v=0$.
See, for example, Fig.\ 7 of Ref.\ \cite{koide_review20}, where the uncertainty always decreases in the early stage of time evolution 
when a stationary fluid is used as the initial condition.
	
\subsection{Inviscid minimum uncertainty} \label{sec:smv}

Let us consider the inviscid limit ($\eta=\mu=0$) in the NSK equation, 
\begin{eqnarray}
(\partial_t + {\bf v} \cdot \nabla ) v^{i} 
= 
-\frac{1}{\uM} \partial_i V + 2\kappa \partial_i \frac{\nabla^2 \sqrt{\rho}}{\sqrt{\rho}} -\frac{1}{\uM\rho} \partial_i P \, .
\label{eqn:madelung}
\end{eqnarray}
This is called the Euler-Korteweg equation.
As was pointed out, the $\kappa$ term  
represents the capillary action in liquid-vapor fluids 
and the gradient of the quantum potential in the Schr\"{o}dinger and Gross-Pitaevskii equations.
Using Eqs.\ (\ref{eqn:mu-state}) and (\ref{eqn:B}), the product of $\sigma^{(2)}_{x^i}$ and $\sigma^{(2)}_{p^j}$ in this case is given by 
\begin{eqnarray}
\sqrt{\sigma^{(2)}_{x^i}} \sqrt{\sigma^{(2)}_{p^j}} 
=
\uM 
\frac{\kappa}{\nu}  \delta_{i j}  
\, . \label{eqn:ucr_nonvis}
\end{eqnarray}
The constant on the right-hand side represents the inviscid minimum uncertainty.

Suppose that the smallest inviscid uncertainty is given by the quantum-mechanical one. 
Then the coefficient $\kappa$ has a lower bound and should satisfy the following inequality 
\begin{eqnarray}
\kappa \ge \frac{\hbar}{2\uM} \nu \, . \label{eqn:lb_kappa}
\end{eqnarray}
When $\kappa$ is fixed, the upper bound of the noise intensity (the diffusion coefficient) is characterized by this inequality. 
This interpretation may sound strange because the uncertainty in our approach comes from the non-differentiability of the stochastic trajectory 
which seems to be enhanced by the increase of the value of $\nu$, as seen from the consistency condition (\ref{eqn:cc}).
Note however that $\kappa$ is proportional to $\nu^2$ as shown by the first equation of Eq.\ (\ref{eqn:coeff}).
Using this, the above inequality is reexpressed as 
\begin{eqnarray}
\nu \ge \frac{\hbar}{4\uM \alpha_B} \, ,
\end{eqnarray}
where $\alpha_B$ is a finite real constant.
As we expected, the uncertainty in Eq.\ (\ref{eqn:ucr_nonvis}) increases as $\nu$ is enhanced.

In the inviscid case, the minimum uncertainty state can be stationary and thus 
satisfies the Fokker-Planck equations (\ref{eqn:ffp}) and (\ref{eqn:bfp}) as was discussed in Sec.\ \ref{sec:qm_limit}.
Moreover, this state satisfies even the Euler-Korteweg equation when we choose 
\begin{eqnarray}
V = \frac{1}{2}\uM \omega^2 x^2 \, , \, \, \, \, \, \, P = C_{pre} \, \rho \, ,
\end{eqnarray}
where $C_{pre} $ is a proportional constant.
Then the stationary solution of the Euler-Korteweg equation is given by 
\begin{eqnarray}
\begin{split}
\rho(x) = \sqrt{\frac{A}{\pi}} e^{-A x^2} \, , \, \,\,\, v(x) = 0\, ,
\end{split}
\end{eqnarray}
where
\begin{eqnarray}
A = \frac{1}{2\kappa} \left\{  \sqrt{\frac{C^2_{pre}}{\uM^2} + 4 \kappa \omega^2}  - \frac{C_{pre}}{\uM} \right\} \, .
\end{eqnarray}
Comparing this solution with Eqs.\ (\ref{eqn:mu-state}) and (\ref{eqn:B}), 
it is easy to see that this stationary state gives the inviscid minimum uncertainty. 
When we use $C_{pre} = 0$ and Eq.\ (\ref{eqn:para_qm}), 
this state agrees with the ground state of the harmonic oscillator in quantum mechanics, which was discussed 
in Sec.\ \ref{sec:qm_limit}

\subsection{Viscous control of minimum uncertainty}

\begin{figure}[t]
\begin{center}
\includegraphics[scale=0.3]{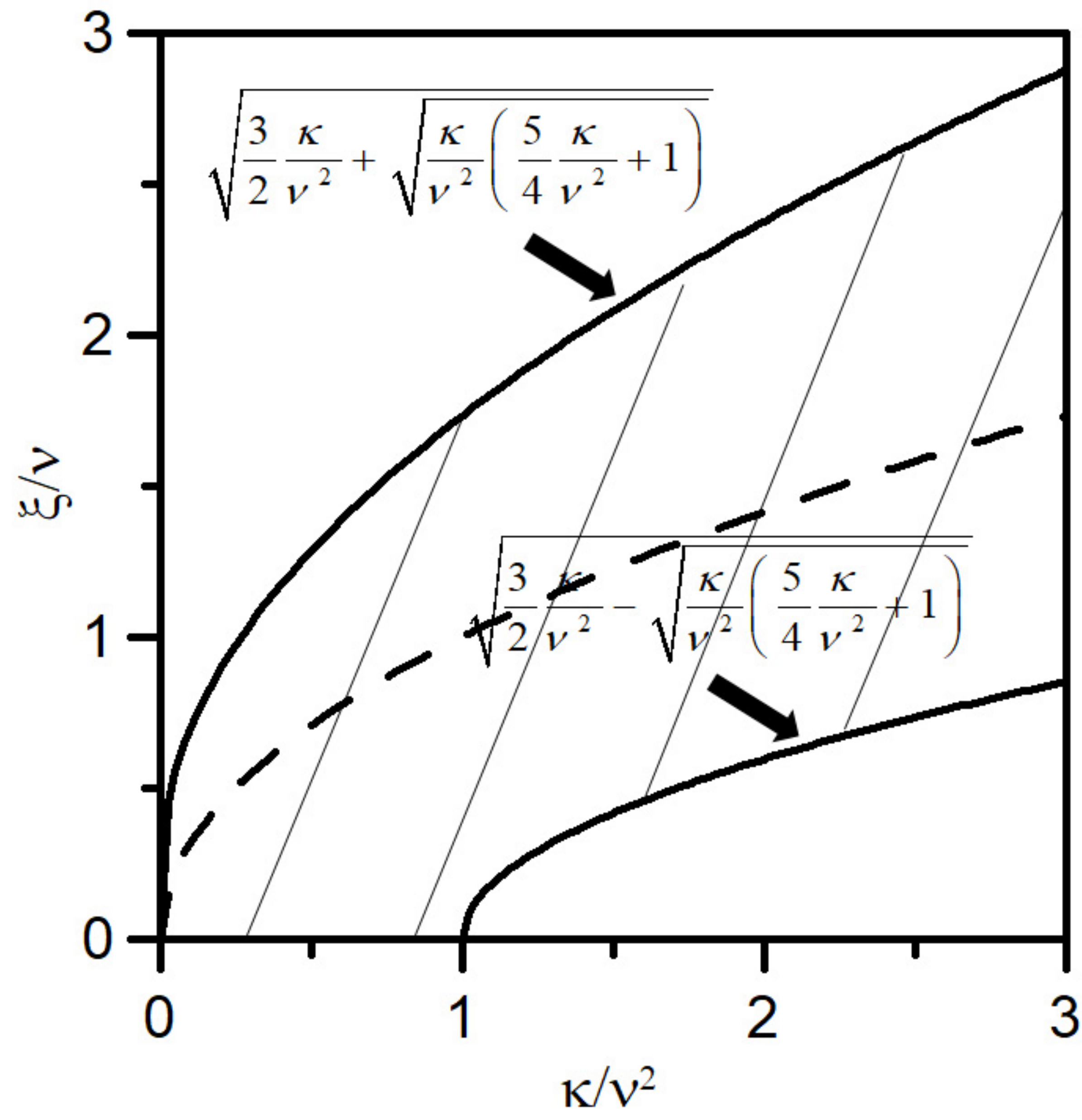}
\end{center}
\caption{
The parameters satisfying the inequality (\ref{eqn:xi_ineq}). 
In the shaded area, the viscous minimum uncertainty is smaller than the inviscid minimum value.
The dashed line represents $\kappa = \xi^2$.}
\label{fig:xi}
\end{figure}

We investigate the effect of viscosity to the inviscid minimum uncertainty defined in Sec.\ \ref{sec:smv}.
The minimum uncertainty obtained from Eqs.\ (\ref{eqn:mu-state}) and (\ref{eqn:B}) is given by 
\begin{eqnarray}
\sqrt{ \sigma^{(2)}_x } \sqrt{ \sigma^{(2)}_p }
= \uM \nu 
 \frac{| (\kappa/\nu^2) - (\xi/\nu)^2 | }{\sqrt{1 + (\xi/\nu)^2}} \, .
\label{eqn:mu_mus}
\end{eqnarray}
For the right-hand side to be smaller than the inviscid minimum, $\uM\kappa/\nu$, 
the kinematic viscosity should satisfy 
\begin{eqnarray}
\xi^*_{min} < \xi < \xi^*_{max} \, , \label{eqn:xi_ineq}
\end{eqnarray}
where
\begin{eqnarray}
\xi^*_{min}
&=&
\left\{
\begin{array}{cl}
 0 &  {\rm for}\, \,  0 \le \kappa <\nu^2 \\
 & \\
\sqrt{\frac{3}{2} \frac{\kappa}{\nu^2} 
- \sqrt{\frac{\kappa}{\nu^2} \left( 1 + \frac{5}{4} \frac{\kappa}{\nu^2} \right)}}
& {\rm for}\, \, \nu^2 \le \kappa
\end{array}
\right. \, , \nonumber \\ 
\xi^*_{max}
&=& 
\sqrt{\frac{3}{2} \frac{\kappa}{\nu^2} 
+ \sqrt{\frac{\kappa}{\nu^2} \left( 1 + \frac{5}{4} \frac{\kappa}{\nu^2} \right)}} \, , \nonumber
\end{eqnarray}
The parameters satisfying these inequalities are shown in Fig.\ \ref{fig:xi}.
The shaded area of the diagram corresponds to the domain where the viscous minimum uncertainty 
is smaller than the inviscid minimum value $\uM \kappa /\nu$.
The uncertainty becomes extremely small around $\kappa = \xi^2$ denoted by the dashed line.
The point $(\kappa/\nu^2,\xi^2/\nu^2) = (1,0)$ on the diagram corresponds to the case of quantum mechanics. 
The NSF equation corresponds to the vertical line of $\kappa/\nu^2 = 0$.

\begin{figure}[t]
\begin{center}
\includegraphics[scale=0.3]{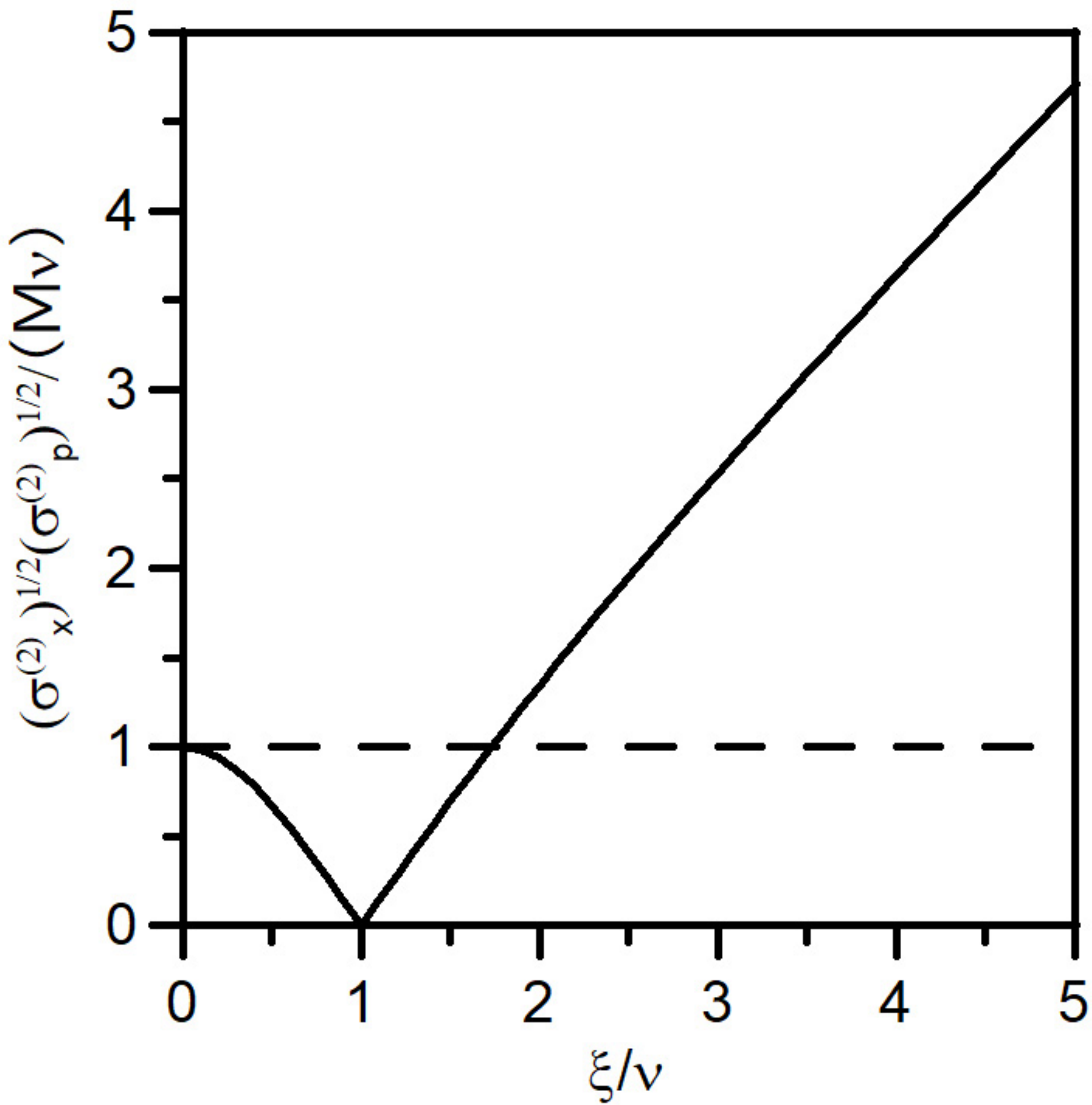}
\end{center}
\caption{The minimum uncertainty for $\kappa=\nu^2$ is plotted as a function of $\xi/\nu$. The dashed line represents the inviscid minimum value $\uM \kappa/\nu = \uM \nu = \hbar/2$ for $\nu = \hbar/(2\uM)$. }
\label{fig:muc}
\end{figure}

As an extreme case, let us consider a weakly interacting quantum many-body system at low temperature.
Then the coefficient $\kappa$ will be given by $\hbar^2/(4\uM^2)$. 
In fact, the Bose-Einstein condensate is approximately described by the Gross-Pitaevskii equation where 
$\kappa = \nu^2 = \hbar^2/(4\uM^2)$.
Suppose that thermal fluctuations induce viscosity 
and the time evolution of the quantum many-body system is described by the NSK equation. 
The coefficient $\kappa$ is a function of temperature and changed from $\hbar^2/(4\uM^2)$ in general. 
We further assume that the temperature dependence in $\kappa$ is weak  
and $\kappa$ is given by $\hbar^2/(4\uM^2)$ at sufficiently low temperature.
In this case, we can treat $\xi$ as a free parameter fixing $\kappa/\nu^2=1$ to investigate 
the behavior of the minimum uncertainty, which is shown in Fig.\ \ref{fig:muc}.
For the sake of comparison, the dashed line represents the inviscid minimum value, 
which agrees with the quantum-mechanical minimum value $\uM \kappa/\nu = \uM \nu = \hbar/2$ in the present parameter set.
The effects induced by the $\kappa$ term and the viscous term cancels each other out and hence we find that 
the product $(\sigma^{(2)}_x)^{1/2} ( \sigma^{(2)}_p)^{1/2}/(\uM \nu)$ can be smaller than $1$ for a sufficiently weak kinematic viscosity satisfying
\begin{eqnarray}
\xi^*_{min} = 0< \xi <  \xi^{*}_{max} = \frac{\sqrt{3}}{2} \frac{\hbar}{\uM}  \, .  \label{eqn:ineq_qm}
\end{eqnarray}

 The viscous minimum uncertainty vanishes when $\xi/\nu = 1$ which correspond to $\kappa = \xi^2$ 
but this choice is forbidden because of the reason discussed in Sec.\ \ref{sec:ideal_limit}. 
For a larger $\xi$, the effect of the viscous term becomes dominant and then the number of the collisions among particles (fluid elements) increases. 
Since the non-differentiability of trajectories is enhanced by the collisions, the viscous minimum uncertainty behaves 
as an increasing function of $\xi$.

\subsection{Lower bound and critical value of viscosity}

For the coefficient $\kappa$, we discussed 
the possible lower bound given by Eq.\ (\ref{eqn:lb_kappa}) for a given $\nu$. 
The similar constraint can exist even for $\xi$.

In relativistic heavy-ion collision physics, 
the behavior of quantum many-body systems is approximately given by a viscous fluid  \cite{hydro_review}. 
The viscous effect is indeed considered to be indispensable 
because it is believed that the shear viscosity cannot be smaller than 
the Kovtun-Son-Starinets (KSS) bound \cite{kss},
\begin{eqnarray}
\frac{\eta}{s} \ge \frac{\hbar}{4\pi k_B} \, , \label{eqn:kss}
\end{eqnarray}
where $s$ is the entropy density. This bound is based on the ansatz of the AdS/CFT correspondence.
Similar lower bounds for the shear viscosity are considered in Ref.\ \cite{gyu}.
Assuming $s \sim  k_B \rho$, 
Eq.\ (\ref{eqn:kss}) reads the inequality for the kinematic viscosity, 
\begin{eqnarray}
\xi \ge \frac{1}{8\pi}\frac{\hbar}{ \uM}  \, .  \label{eqn:kss2}
\end{eqnarray}

Let us consider the relation between the viscous minimum uncertainty 
and the KSS bound in the system considered in Fig.\ \ref{fig:muc}.
To have a minimum value smaller than the inviscid one $\hbar/2$, 
the kinematic viscosity should be smaller than the critical value $\xi^*_{max}$ defined in Eq.\ (\ref{eqn:ineq_qm}). 
Comparing Eq.\ (\ref{eqn:kss2}) with Eq.\ (\ref{eqn:ineq_qm}), 
we find that $\xi^*_{max}$ is larger than the KSS bound 
and thus the viscous effect can induce the minimum uncertainty smaller than $\hbar/2$ in principle at least. 
However, the difference between the critical value and the KSS bound is only slight 
and thus we cannot decide the precise order of these quantities in the present rough estimation.
The KSS bound may indicate that 
there exists a fundamental mechanism in quantum physics which does not permit 
the improvement of uncertainty beyond the quantum-mechanical minimum value $\hbar/2$ by viscosity.

\section{Concluding remarks}
\label{sec:conclusion}

The viscous minimum uncertainty state of the fluid described by the Navier-Stokes-Korteweg equation was derived.
This state has a Gaussian particle distribution and thus is regarded 
as the generalization of the coherent state of quantum mechanics. 
The velocity field of this state exhibits the linear-position dependence and the inclination 
is characterized by the shear viscosity.
Such a linear dependence in the velocity field is often observed in the expanding fluid described by the Navier-Stokes-Fourier equation.
The corresponding uncertainty is controlled by the shear viscosity and can be smaller 
than the inviscid minimum value when the shear viscosity is smaller than a critical value.
The parameter set to satisfy this condition distributes zonally on the diagram of the transport coefficients 
$\kappa$ and $\xi$ as is shown in Fig.\ \ref{fig:xi}.

The existence of such a parameter set requires special attention
because the shear viscosity can have a minimum value. 
If this minimum is given by the Kovtun-Son-Starinets bound, 
we found that the order of the KSS bound is similar to that of the critical value of the shear viscosity. 
This may suggest that the lower bound of viscosity appears so that 
viscosity does not improve uncertainty beyond the quantum-mechanical minimum value $\hbar/2$.

The shear and bulk viscosities in Eq.\ (\ref{eqn:nsk}) are given by the same formulae 
as those of the NSF equation.
As is well-known, these coefficients are determined by the Green-Kubo-Nakano (GKN) formula \cite{zwanzig,koide_tra1,koide_tra2,koide_tra3}.  See, for example, the discussion around Eq.\ (26) of Ref.\ \cite{koide_tra1}. The shear viscosity $\eta$ is given by Eq.\ (27). 
To obtain this expression, we linearize the NSF equation and take the low wave number limit 
(${\bf k} \rightarrow 0$). 
Applying this procedure to Eq.\ (\ref{eqn:nsk}), the contribution from the $\kappa$ term disappears and thus Eq. (1) coincides with the NSF equation. 
Therefore the GKN formula of the NSF equation is applicable to determine the coefficients in Eq.\ (\ref{eqn:nsk}).
The corresponding formula of the coefficient $\kappa$ is however not yet known.
The formulation developed in Refs.\ \cite{koide_tra1,koide_tra2,koide_tra3} is applicable not only to 
classical many-body systems but also to quantum many-body systems.
Thus, if such a formula is found, the coefficient $\kappa$ is calculated from quantum mechanics.
Differently from the coefficients of irreversible currents, the $\kappa$ term does not violate the time-reversal symmetry in the NSK equation 
and thus will be characterized by the real part of the retarded Green's function of microscopic currents.

In the standard formulation of quantum mechanics, 
the non-commutativity of operators leads to the uncertainty relation, 
while the same property is reproduced from the non-differentiability of particle trajectories in the present approach.
The operator formalism is established in various applications of quantum mechanics and thus 
one may wonder about the significance of the alternative interpretation for the uncertainty relation.
The advantage of the present approach is its applicability to generalized coordinate systems. 
For example, the angle variable and the angular momentum form a pair of canonical variables in polar coordinates, 
but the corresponding operator representations are not established  
because there is no self-adjoint multiplicative operator which satisfies the periodicity and the canonical commutation relation simultaneously. See Ref.\ \cite{koide20-1} and references therein. 
Therefore, in the discussion of the angular uncertainty relation, the angle operator is introduced exclusively by altering one of those conditions.
By contrast, the present approach is applicable to quantize generalized coordinate systems without introducing additional condition \cite{koide19} and the uncertainty relation in generalized coordinates is obtained 
without any difficulty \cite{koide20-1}.

It is known that thermal fluctuations are enhanced in low dimensions $(D <3)$ and 
such strong fluctuations can trigger modification of Eq.\ (\ref{eqn:nsk}) \cite{ernst,kovtun}.
In our approach, this difference of fluctuations will be taken into account through Brownian motions.
In one and two dimensions, Brownian motion is recurrent: a Brownian particle comes back to an initial position at some time or other.
However, in higher dimensions ($D \ge 3$), the trajectory is not recurrent. See, for example, Ref.\ \cite{ezawa} and references therein.
Thus the difference of fluctuations in low and high dimensions will be investigated 
through the comparison of the uncertainty relations.
For example, if the uncertainty in low dimensions is not qualitatively different from the one 
in high dimensions, 
it may be the signature of the incompatibility of Eq. (\ref{eqn:nsk}) in low dimensions.

The viscous uncertainty characterizes the motion of fluid elements.
The fluid element is an abstract volume element and thus its direct observation will be difficult.
However, the descriptions based on hydrodynamic models sometimes depend on 
the motions of fluid elements and thus the existence of 
the viscous uncertainty triggers the modification of the descriptions.
Physics in relativistic heavy-ion collisions is one example \cite{hydro_review}. 
The vacuum is excited by high-energy nucleus collisions and the behavior of the excited vacuum is approximately described by viscous hydrodynamics. 
The experimentally observed particles, called hadrons, are assumed to be produced by the thermal radiation from each fluid element of the viscous fluid. 
It is known that this hydrodynamic model explains experimental data very well. 
In this model, we assume that the fluid elements pass along the streamline of the viscous fluid, but 
such a view is too simple. 
Our result shows that the currents of the fluid elements fluctuate around streamlines of the viscous fluid and this fluctuation 
is characterized by $\sigma^{(2)}_p$.
Moreover 
the behavior of $\sigma^{(2)}_p$ is restricted by that of $\sigma^{(2)}_x$ which can reflect the inhomogeneity of the matter distribution.
See also Fig.\ 5 in Ref.\ \cite{koide_review20}.
Because of the lack of the above mentioned effect, 
the standard hydrodynamic model may underestimate the effect of 
the spatial inhomogeneity of the excited vacuum and hence the anisotropy of the hadron production.
A more quantitative analysis is left as a future work.

\vspace*{1cm}
The author thanks J.-P.\ Gazeau and T.\ Kodama for fruitful discussions and comments, 
and acknowledges the financial support by CNPq (303468/2018-1).
A part of the work was developed under the project INCT-FNA Proc.\ No.\ 464898/2014-5.

\end{document}